\documentstyle[11pt,IAU207_pasp,twoside,psfig]{article}
\newcommand{\etal}{{et al.\ }}

\newcommand{\gta}{\stackrel{>}{\scriptstyle\sim}}

\def\arg#1{{\it#1\/}}

\def\edcomment#1{\iffalse\marginpar{\raggedright\sl#1\/}\else\relax\fi}
\reversemarginpar

\begin{document}
\title{Predicting Metallicities and Color Distributions \\
for Secondary GCs Forming in Spiral Galaxy Mergers at Various Redshifts}
 \author{Uta Fritze -- v. Alvensleben}
\affil{Universit\"atssternwarte G\"ottingen, Geismarlandstr. 11, 
37083 G\"ottingen, Germany}

\begin{abstract}
In a 1$^{\rm st}$ step I present results from our new set of evolutionary synthesis models for Simple (= single burst) Stellar Populations ({\bf SSP}s) of various metallicities, and in a 2$^{\rm nd}$ step I combine these results with the information we have about the redshift evolution of spiral galaxies' ISM abundances. The aim is to provide a grid of color and luminosity distibutions of any secondary GC population formed at some time in the past in those mergers for comparison with observations. 
\end{abstract}

\section{Introduction}
F. Schweizer's contribution ({\sl this conf.}) has shown that in the local Universe, we can witness the formation of -- sometimes populous -- star cluster systems in the powerful starbursts accompanying gas-rich spiral galaxy mergers. Many of these clusters are compact and massive enough to be Globular Clusters ({\bf GC}s) able to survive for many Gyr. Formed from pre-enriched gas they will stand out in metallicity for all times, even when their colors and luminosities will have become inconspicuous. Spirals were more gas-rich in the past and mergers were more frequent. The origin of elliptical galaxies is still a matter of debate. Were their stars formed ``all at once'' in the early Universe -- either in a monolithic collapse scenario or in 
some ``coordinated effort'' in building blocks that later merged stellardynamically -- or do they have significant populations of stars formed in one major spiral-spiral merger or in a series of hierarchical mergers that still involved gas to power star formation. Observations of metallicity and color distribution of GC systems may tell us more about a galaxy's formation history than its integrated light (cf. S. Zepf, {\sl this conf.}). Spectroscopy of reasonable numbers of GCs in elliptical or S0 galaxies reaches its limits at Virgo cluster distances and has to cope with crowding and the galaxy background. GC color distributions from HST observations are becoming available for many galaxies (e.g. Gebhardt \& Kissler -- Patig 1999). 

In a 1$^{\rm st}$ step I present results from our new set of evolutionary synthesis models for Simple (= single burst) Stellar Populations ({\bf SSP}s) of various metallicities, and in a 2$^{\rm nd}$ step I combine these results with the information we have about the redshift evolution of spiral galaxies' ISM abundances. The aim is to provide a grid of color and luminosity distributions of any secondary GC population formed at some time in the past for comparison with observations.  

\section{Evolutionary Synthesis of SSPs}
Using Padova isochrones including the thermal pulsing AGB phase ({\bf TP-AGB}) we obtain the time evolution of spectra (90 \AA -- 160 ${\rm \mu m}$), colors, luminosities (U ... K), and stellar mass loss from ages $1 \cdot 10^8 - 15 \cdot 10^9$ yr for SSPs of 5 different metallicities [Fe/H] $=~1.7,~-0.7,~-0.4,~0,~+0.4$ (Schulz, Fritze -- v. A., Fricke 2001). Inclusion of the TP-AGB phase in the stellar evolutionary tracks is very important for the (V-I) and (V-K) colors at ages $\gta 6 \cdot 10^7$ yr, more so for higher metallicities than for lower ones. For ${\rm Z_{\odot}}$ it makes colors redder by $> 0.3$ mag in (V-I) and $> 1$ mag in (V-K) at ages $1 \cdot 10^8 - 1 \cdot 10^9$ yr. This means that ages determined for young star clusters from HST (V-I) colors and models without TP-AGB are strongly overestimated.

Without going into any details, we only recall here that -- compared to solar metallicity -- SSPs at lower metallicities are brighter in UBVRI and fainter in K, are bluer and have lower mass loss, lower M/L$_{\rm B,V}$ and higher M/L$_{\rm K}$ values all through their time evolution (Fritze -- v. A. 2000). Models are shown to not only correctly reproduce the observed GC colors after a Hubble time but also the empirical calibrations like (B-V) or (V-I) vs [Fe/H] for Milky Way ({\bf MW}) GCs. 
Models also show how strongly those calibrations change as clusters get younger (cf. Schulz \etal 2001 for details). 

In our 1$^{\rm st}$ step, we compare the time evolution of GC ($=$ SSP) model colors of various metallicities with the observed mean colors of MW halo GCs from Barmby \etal's (2000) compilation. We find that 
low metallicity clusters ([Fe/H]$= -1.7$) start out bluer than MW GCs with ${\rm \langle [Fe/H]\rangle = -1.35}$ and come close to their mean colors at ages $> 5$ Gyr both in (B-V) and in (V-I). In (V-K) these low metallicity GCs are close to the MW halo clusters at all ages. At intermediate metallicity [Fe/H]$=-0.4$, clusters also start out bluer than MW GCs in (B-V) and (V-I), they cross the MW GC colors around ages of 2 Gyr and become redder thereafter. In (V-K) they are slightly redder than MW GCs from the very beginning. High metallicity clusters [Fe/H]$=+0.4$ already have mean MW GC colors at ages around 1 Gyr in (B-V) and (V-I) and get considerably redder thereafter. In (V-K) they already are redder than MW GCs at 0.3 Gyr. 

Note that certain combinations in age and metallicity, that are indistinguishable in optical colors due to the well-known age -- metallicity degeneracy, do visibly split up in (V-K). 

{\bf We conclude} that an observed unimodal GC color distribution need not necessarily imply that all GCs have the same age and metallicity. Two (or more) GC populations may hide in one peak. In case of an observed bimodal GC color distribution it is not {\sl a priori} clear if a red or blue peak is due to an older/younger or metal-rich/metal-poor cluster population. Individual cluster spectroscopy is required to clarify the situation.  

In Fritze -- v. A. (2001) we also study the luminosity functions ({\bf LF}s) of these secondary GCs in their time evolution. We assume a mass function similar to that of the MW GCs and gauge the luminosity of an SSP with [Fe/H]=-1.35 to their mean ${\rm \langle M_{V_o}\rangle = -7.3}$ to see for which combinations of age and metallicity the LF of a model GC population could be distinguished from the observed MW GC LF. Depending on the specific combinations of ages and metallicities all combinations are seen to appear for two GC populations: uni- and bimodal color distributions with uni- and bimodal LFs.

\section{Color Distributions and Luminosity Functions of GCs Forming in Spirals-Spiral Mergers at Various Redshifts}
In evolutionary synthesis models galaxies of various spectral types are para\-me\-trised by their respective star formation ({\bf SF}) histories. Starting from an initial gas cloud and specifying a SF history and an IMF, our models simultaneously yield the spectrophotometric evolution of the stellar population and the chemical evolution in terms of ISM abundances, both as a function of time and, for any kind of cosmological model, also as a function of redshift. Following both aspects of galaxy evolution simultaneously allows to describe the evolution in a chemically consistent way, i.e. to monitor the initial (= gas phase) abundance at which any star is born and follow it on a track and with yields appropriate for its metallicity (Fritze -- v. A. 1999). This gives results in good agreement with observed stellar metallicity distributions, template spectra, and HII region abundances of local galaxy types as well as with the observed redshift evolution of luminosities and colors (M\"oller \etal 1999). Comparison with precise individual element abundances measured on high resolution spectra of Damped Ly$\alpha$ absorbers for a series of elements (Fe, Si, Zn, Cr, Ni, S, Al, Mn) shows good agreement with the redshift evolution of the absolute abundances in our spiral models over a redshift range from ${\rm z>4}$ to ${\rm z\sim 0.4}$. By ${\rm z=0}$ models directly meet observed characteristic HII region abundances. The observed weak redshift evolution of abundances and the large scatter at any redshift are seen to be a consequence of the SF timescales in spirals and their differences from early to late-type spirals (cf. Lindner \etal 1999 for details). Optical properties of the models are in agreement with the large number of non-detections and with the properties of the few detected Damped Ly$\alpha$ absorbers (Fritze -- v. A. \etal 1999). 
Hence, we believe that we reasonably understand the redshift evolution of spiral galaxies' ISM abundances over large lookback times, and in the following, I will use these to investigate the metallicities of hypothetic secondary populations of GCs forming from this ISM in spiral -- spiral mergers at various redshifts. Combining with SSP models for the respective metallicities, the time evolution of colors and characteristic luminosities of these secondary GCs are obtained. 

In Fritze -- v. A. \& Gerhard (1994) we predicted abundances for the young star clusters in the Sc -- Sc merger remnant NGC 7252 to be ${\rm \gta \frac{1}{2}~Z_{\odot}}$. Spectroscopy of the brightest clusters, i.e. those formed towards the end of the extended starburst, yielded abundances ${\rm Z \sim Z_{\odot}}$ with some indication of $\alpha$-element enhancement from the burst itself (Schweizer \& Seitzer 1993, 1998, Fritze -- v. A. \& Burkert 1995) in good agreement with model predictions. 

The redshift evolution of ISM abundances for various spiral types shows that only a limited range of combinations between metallicity and age of secondary GCs is expected to result from mergers of normal spirals. Despite a significant amount of scatter introduced by the metallicity differences among different spiral types at fixed redshift, a broad cosmic age -- metallicity relation is found. The ISM metallicity in spirals increases slowly with decreasing redshift. Hence secondary GCs formed out of this gas are predicted to be the more metal-rich the later the merger, i.e. the younger the GCs are at the present time. SSP models have shown that metallicity effects on the evolution of a GC depend both on wavelength/color and on age. Hence, it is the interplay of age and metallicity that determines the color and luminosity evolution. 

\begin{figure}
\centerline{\vbox{
\psfig{figure=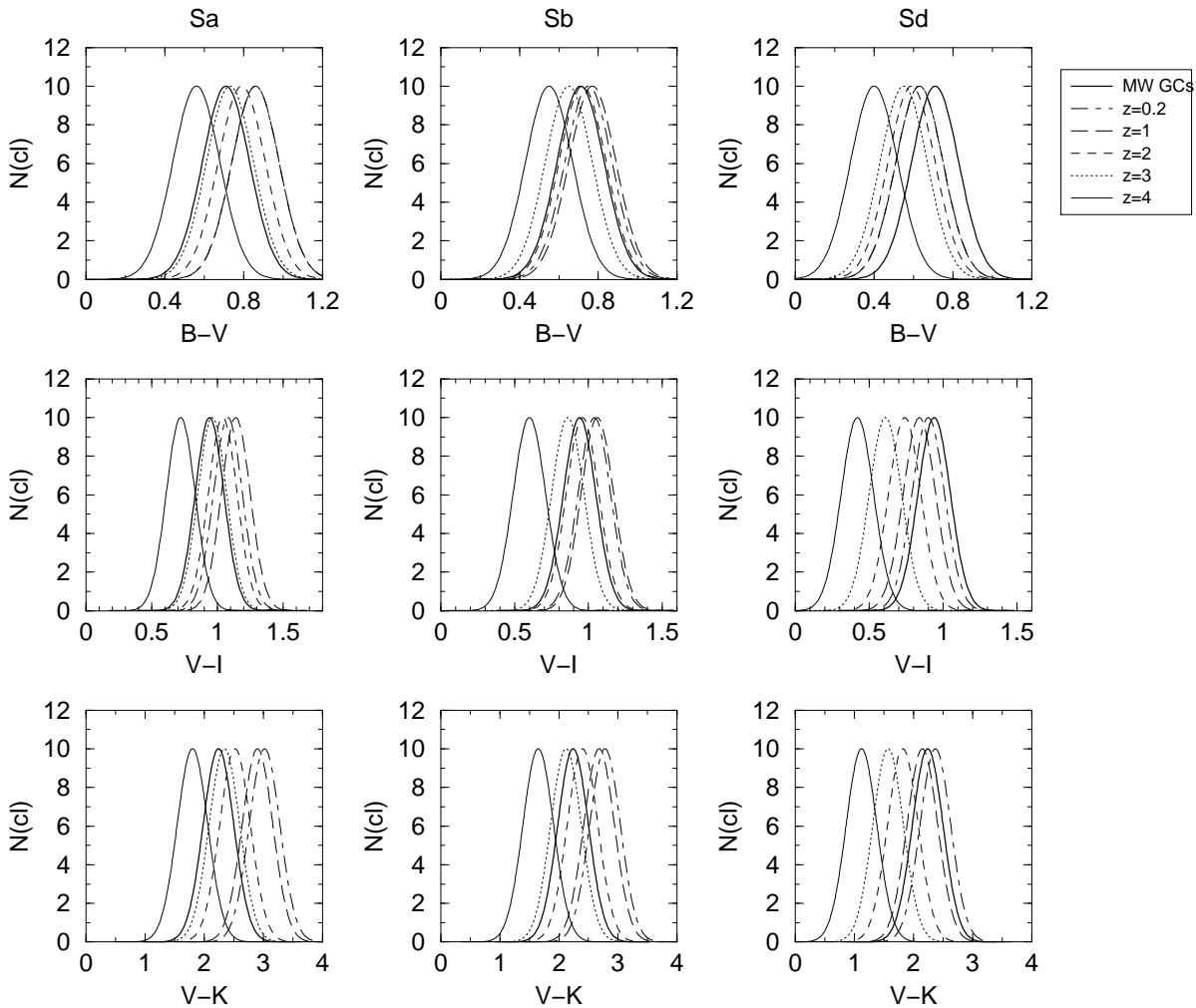,width=13cm}
}}
\caption{Color distributions for secondary GC populations formed in mergers of differnt spiral types at various redshifts. 
First column: Sa -- Sa mergers, second column: Sb -- Sb mergers, 
third column: Sd -- Sd mergers. The Milky Way halo GC color distributions are taken as a reference for the primary GCs.}
\end{figure}

In Fig. 1 we present the color distributions as predicted for today for secondary GCs that formed in spiral -- spiral mergers at various redshifts. We also plot the observed color distributions of MW halo GCs which we take as a reference for the primary GC population in our mergers. We assume the same intrinsic widths for the color distributions of primary and secondary cluster systems and plot a situation where both populations contain comparable numbers of clusters. In reality, however, the two peaks may have different heights according to the relative numbers of primary and secondary GCs. The number of secondary GCs will depend on the amount of gas available and on the cluster formation efficiency in the burst. A 2$^{\rm nd}$ peak is easiest to detect if comparable in height to the primary. Note that many -- if not most -- of the observed GC systems show a peak in their (V-I) distribution at or close to that of the MW halo GCs, suggesting a certain uniformity of primary GC systems. 

Fig. 1 shows that the color distributions of secondary GCs are bluer than those of the MW GCs for all very early mergers at ${\rm z\sim 4}$ and Sd-Sd mergers at ${\rm z \geq 2}$, they are hidden in the primary GC color distributions for Sa-Sa mergers at ${\rm z\sim 3}$, Sb-Sb mergers at ${\rm 3 \geq z \geq 2}$, and Sd-Sd mergers at ${\rm z\sim 0.2}$. Color distributions of secondary GCs finally are redder than those of MW GCs for Sa-Sa mergers at ${\rm z < 2}$, Sb-Sb mergers at ${\rm z\leq 1}$. 

It is very intriguing and a result of the restricted range of age -- metallicity combinations in spiral mergers that LFs in all cases -- even in those with clearly bimodal color distributions -- are unimodal and narrow, thus ``saving'' this method of distance determination (cf. Fritze -- v. A. 2001). Observed GC LFs, as well, are unimodal in several systems with bimodal color distributions (e.g. M87, NGC 4472, ...).

\section{Outlook} 
These first and very simplified models need to be extended to mergers among galaxies of different types, including gas-rich dwarfs. Observationally it is very important to have imaging in more than two bands, and in particular in the NIR, and to obtain spectra of reasonable numbers of GCs in both kinds of systems, those with bimodal and rich unimodal color distributions, to reveal if they contain different metallicity subpopulations that -- for specific ages -- may perfectly hide within one peak of the color distribution. 


\section*{Discussion}

\noindent{\it Leo Girardi:\, } Let me add a small comment to your important point about the necessity 
to have TP-AGB stars in the SSP models: fortunately, nowadays most models include the TP-AGB in 
a reasonable way (as for instance yours, Padova, and Bruzual \& Charlot's), but we should also 
take care that there are other models where the TP-AGB is included in a way that simply does 
not make sense from the point of view of stellar evolution theory.\\

\noindent{\it Jan Palous:\, } How do you mix the chemical elements in a merger. Do you assume that the violent relaxation mixes the ISM and smears out the metallicity gradients?\\

\noindent{\it Uta Fritze -- v. A.:\, } Our models do not include any dynamics or spatial resolution. We assume that after the lifetime of a star all the H, He, and heavy elements it gives back are instantaneously and perfectly mixed with the ISM. In mergers with strong global starbursts able to form new GCs this approximation may not be too bad due to the efficiency of stirring at work. 

\bigskip\noindent
{\bf Acknowledgement:} I gratefully acknowledge travel support from the DFG (Fr 916/9-1) and from the organisers.
  

\begin{references}

\reference Barmby, P., Huchra, J. P., Brodie, J. P., D. A. Forbes, L. L. Schroeder \& C. J. Grillmair 2000, \aj \ 121, 1482
 
\reference Fritze -- v. Alvensleben, U. 1999, in ASP Conf. Ser. 192, Spectrophotometric Dating of Stars and Galaxies, eds. 
I. Hubeny, S. R. Heap, R. H. Cornett, 273

\reference Fritze -- v. Alvensleben, U. 2000, in ASP Conf. Ser. 211, Massive Stellar Clusters, eds. A. Lan\,con \& C. M. Boily, 3

\reference Fritze -- v. Alvensleben, U. 2001, \aap \ {\sl in prep.}

\reference Fritze -- v. Alvensleben, U. \& Burkert, A. 1995, \aap \ 300, 58

\reference Fritze -- v. Alvensleben, U. \& Gerhard, O. E. 1994, \aap \ 285, 751 + 775


\reference Fritze -- v. Alvensleben, U. 1999, in Chemical Evolution from Zero to high Redshift, eds. J. R. Walsh, M. R. Rosa, Springer 1999, p. 256 

\reference Gebhardt, K. \& Kissler -- Patig, M. 1999, \aj \ 118, 1526 


\reference Lindner, U., Fritze -- v. Alvensleben, U. \& Fricke, K. J. 1999, \aap \ 341, 709

\reference M\"oller, C. S., Fritze -- v. Alvensleben, U. \& Fricke, K. J. 1999, IAU Symp. 183, 158

\reference Schulz, J., Fritze -- v. Alvensleben, U. \& Fricke, K. J. 2001, \aap \ {\sl in prep.}

\reference Schweizer, F. \& Seitzer, P., 1993 \apj \ 417, L29

\reference Schweizer, F. \& Seitzer, P. 1998, \aj \ 116, 2206


\end{references}
\end{document}